\def\la{\mathrel{\mathpalette\fun <}}
\def\ga{\mathrel{\mathpalette\fun >}}
\def\be{\begin{equation}}
\def\ee{\end{equation}}
\def\ba{\begin{eqnarray}}
\def\ea{\end{eqnarray}}

\documentstyle[11pt,paspconf,epsf]{article}

\begin{document}

\title{A Model-Independent Photometric Redshift Estimator}

\author{Yun Wang, Neta Bahcall, and Edwin L. Turner}
\affil{Princeton University Observatory, Peyton Hall, Princeton, NJ 08544}

\begin{abstract}

We derive a simple empirical photometric redshift estimator using a 
training set of galaxies with multiband photometry and measured redshifts
in the Hubble Deep Field (HDF).
This estimator is model-independent; it does not use spectral templates.
The dispersion between the estimated redshifts and the spectroscopically
measured ones is small; the dispersions range from $\sigma_z\simeq 0.03$
to 0.1 for $z\la 2$ galaxies, and from $\sigma_z\simeq 0.14$ to 0.25
for $z\ga 2$ galaxies. The predictions provided by
our empirical redshift estimator agree well with
recently measured galaxy redshifts. We illustrate how
our empirical redshift estimator can be modified to include
flat spectrum galaxies with $1.4\la z \la 2$.

\end{abstract}

% Keywords should be included, but they are not printed in the hardcopy.


\section{Introduction}

Our aim is to derive a simple but accurate empirical photometric redshift 
estimator which is model-independent. We do not use spectral templates;
we only use a training set of galaxies with multiband photometry and 
measured redshifts. We have applied our method
to the Hubble Deep Field (HDF) (Williams et al. 1996) 
and derived a catalog of estimated
redshifts for 848 HDF galaxies with $I<27$ and with measured fluxes
in $U$, $B$, $V$, and $I$ (Wang, Bahcall, \& Turner 1998).

To derive formulae for the estimated redshifts, we first divide the
galaxy redshift sample (the training set) into regions of high and low 
redshifts ($z\ga 2$ and $z\la 2$) based on empirical color cuts.
We then divide both regions into color ranges to minimize dispersion.
The physical motivation for this latter step is to reflect color shifts
with $z$ for different type galaxies.
We then find the best linear fit between redshift and colors
for each color range, 
\be
z_a=c_1+c_2(U-B)+c_3(B-V)+c_4(V-I),
\ee
where $c_i$ ($i=1,2,3,4$) are constants.
Here we have assumed that photometry has been done in four bands, $U$, $B$, 
$V$, and $I$. The dispersion ($\sigma_z$) between our estimated redshift, 
$z_a$, and the measured spectroscopic redshift, $z$, 
is calculated using the jack knife method.
%To find the dispersion via the jack knife 
For a sample with $N$ data points, 
we make $N$ subsamples, each omitting one data point.
We then carry out linear fits $N$ times on the $N-1$ data point subsamples.
We use each fit to make a prediction for the omitted data
point; the rms of the true values of the omitted 
points about the fit predictions, scaled by
$[(N-1)/N]^{1/2}$, is the rms dispersion of our fit.
The estimated dispersions of our formulae are therefore quite robust.

The error in $z_a$ due to photometric errors is
\be
{\Delta}z_a=\left[(c_2{\Delta}U)^2+|(c_3-c_2){\Delta}B|^2+
|(c_4-c_3){\Delta}V|^2+(c_4{\Delta}I)^2\right]^{1/2}.
\ee

\section{Application to the HDF}

We used 82 galaxies with measured spectroscopic redshifts $z\simeq 0.1-3.5$
in the HDF as our training set
(Cohen et al. 1996; Hogg et al. 1998; Lowenthal et al. 1997; 
Phillips et al 1997; Steidel et al. 1996).
We did not use all 90 (excluding three $z>2$ galaxies with 
erroneous/uncertain $z$'s) HDF galaxies with measured $z$'s. The eight 
outlying galaxies have $\sigma_z=0.45$,
mostly near the boundaries of the color ranges; 
these galaxies were not used in deriving our redshift estimator.

We found that for $z\ga 2$, the galaxies satisfy at least one of
the following three color selection criteria:
\be
 U\ge 25.66, \,U-B\ge 0.91,\, B-V\le 1.37,\, V-I\le 0.5;
\ee
\be
 I>23.5,\, U-B>2.2;
\ee
\be
 I>23.5, \,B-V>2.2,\, U-B>-0.5.
\ee
The galaxies with $z<2$ in the training set do not satisfy any of the above
relations, suggesting that $z<2$ galaxies generally fall outside these
color-magnitude regions.		

We divide $z<2$ galaxies into three color ranges (cr), and determine
the best-fit redshift estimate formula and dispersion for each color range:

\noindent{1. cr=1: $(U-B)<(B-V)-0.1$ (28 galaxies)}
\be
\label{eq:Lz3,1}
z_a=0.4111-0.1852(U-B)-0.3062(B-V)+0.7301(V-I),
\hskip 0.7cm \sigma_z=0.034;
\ee
2. cr=2: $(U-B)\ge (B-V)-0.1>(V-I)$ (21 galaxies)
\be
\label{eq:Lz3,2}
z_a=0.163-0.171(U-B)+0.340(B-V)+0.194(V-I),
\hskip 1cm \sigma_z=0.095;
\ee
3. cr=3: $(U-B)\ge (B-V)-0.1\le(V-I)$ (19 galaxies)
\be
\label{eq:Lz3,3}
z_a=1.126+0.480(U-B)-0.513(B-V)-0.250(V-I),
\hskip 1cm \sigma_z=0.097.
\ee

We divide $z\ge 2$ galaxies into two color ranges, and determine
their redshift estimate formulas and dispersions:

\noindent{1. cr=4: $(B-V)-0.5>(V-I)$ (8 galaxies; $z\ga 3$)}
\be
\label{eq:Hz2,1}
z_a=2.37+0.02(U-B)+1.61(B-V)-2.47(V-I),
\hskip 1cm \sigma_z=0.14;
\ee
2. cr=5: $(B-V)-0.5\le (V-I)$ (6 galaxies; $z\la 3$)
\be
\label{eq:Hz2,2}
z_a=2.18+0.10(U-B)+0.20(B-V)+0.75(V-I),
\hskip 1cm \sigma_z=0.25
\ee

Fig.1 shows our estimated redshift $z_a$ (given by 
Eqs.(\ref{eq:Lz3,1})-(\ref{eq:Hz2,2}))
versus the spectroscopic redshift $z$ for 90 HDF galaxies.
%The galaxies with known measurement errors in $UBVI$ are plotted with error 
%bars in $z_a$.
%Error bars without points represent the eight galaxies not used in fitting.
\begin{figure}
\plotone{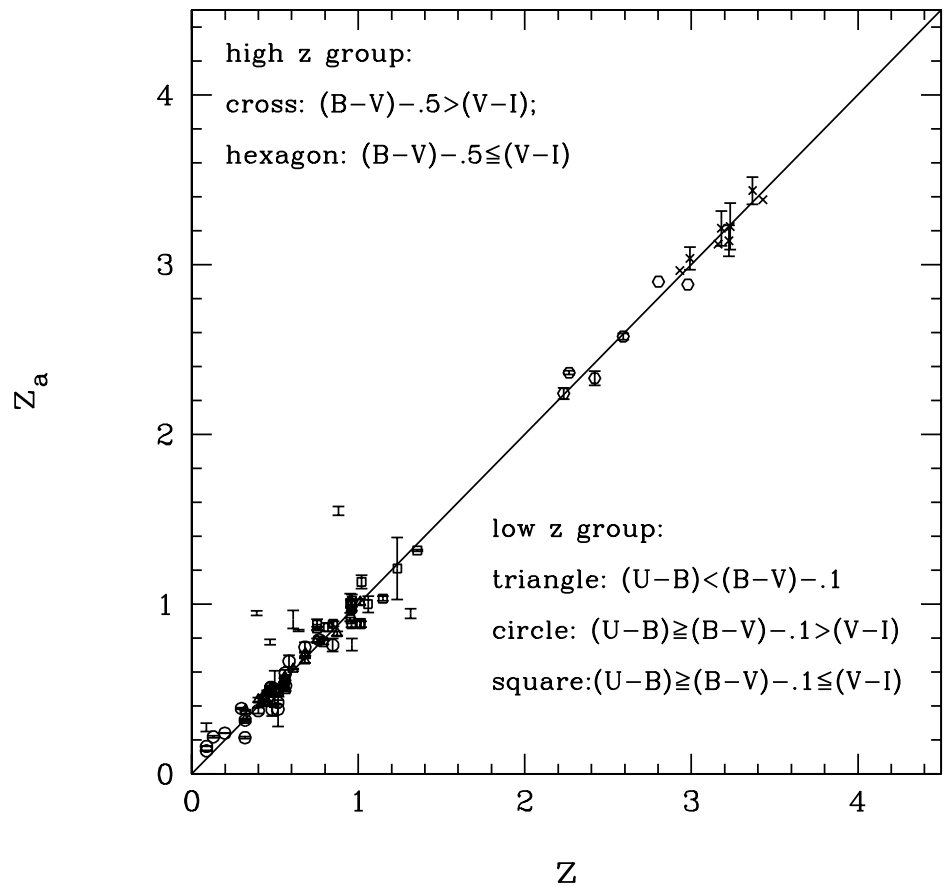} 
\caption{Our empirically estimated redshift $z_a$ (given by 
Eqs.(\ref{eq:Lz3,1})-(\ref{eq:Hz2,2}))
versus the spectroscopic redshift $z$ for 90 HDF galaxies. 
The galaxies with known measurement errors in $UBVI$ are plotted with error 
bars in $z_a$.
Error bars without points represent the eight galaxies not used in fitting.} 
\end{figure}

Our empirical photometric redshift estimator is consistent
with template-based photometric redshift estimators where calibrating
redshift samples are available. Fig.2(a) compares our empirically estimated 
redshift $z_a$ (given by Eqs.(\ref{eq:Lz3,1})-(\ref{eq:Hz2,2}))
with the template-fitting photometric redshift $z_{temp}$ of
Sawicki et. al. (1997), for 848 galaxies in the HDF
with $I<27$ and measured $UBVI$. 
The symbols are the same as in Fig.1.
The solid diagonal line indicates $z_a=z_{temp}$;
the dotted lines mark the region $|z_a-z_{temp}|\leq 0.5$.
About 90\% of the galaxies fall within the dotted lines.
\begin{figure}
\plottwo{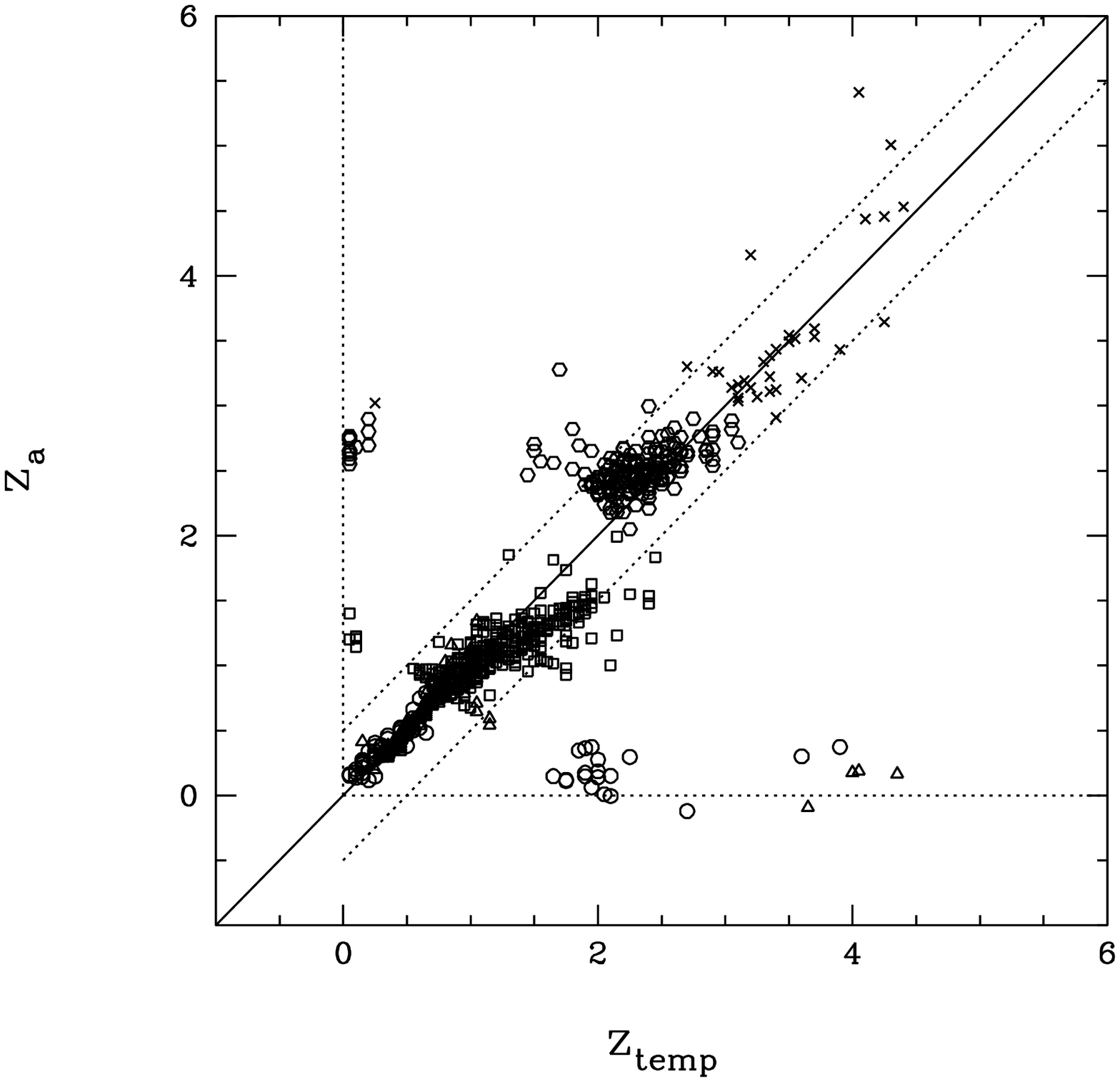}{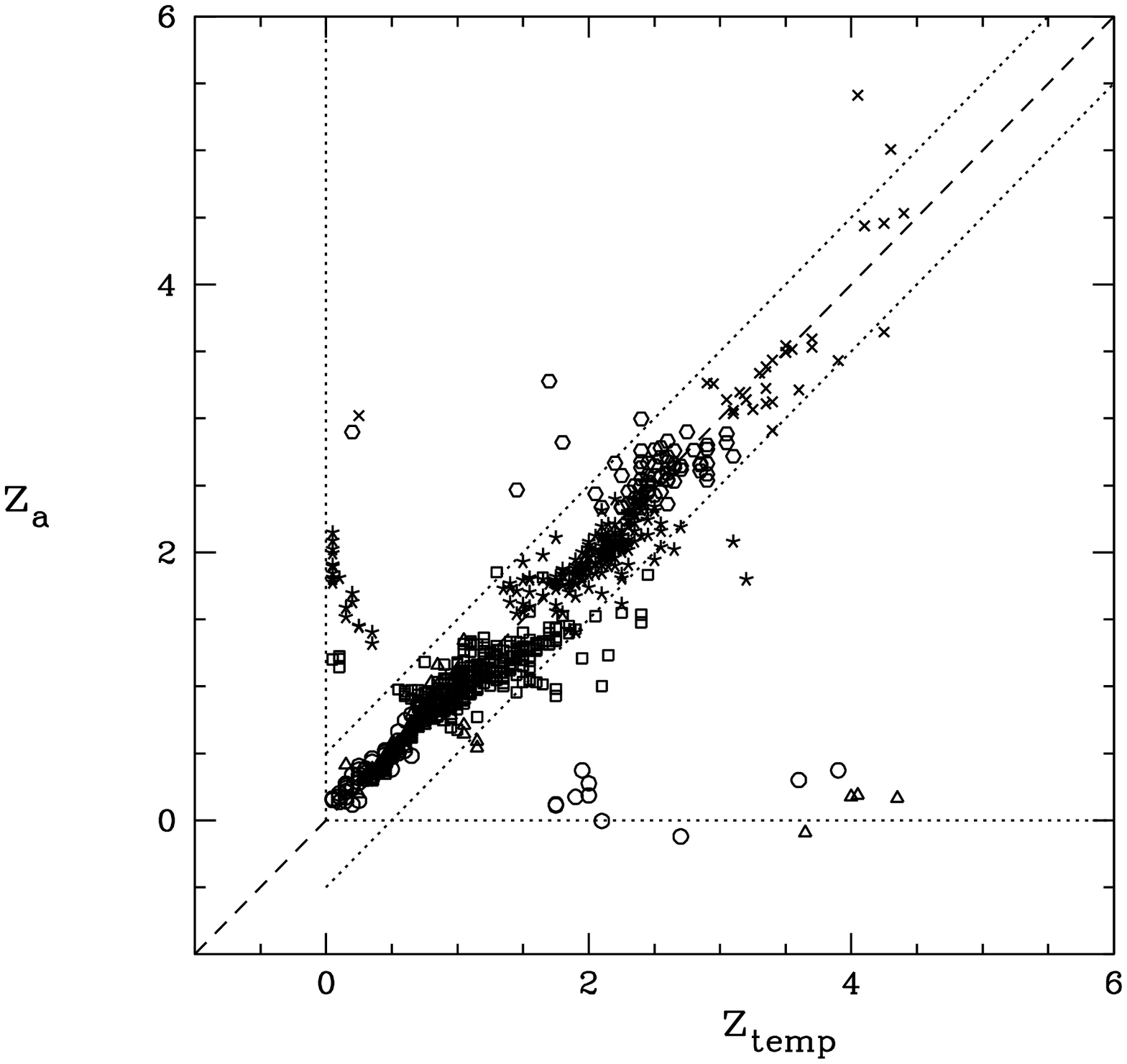}
\caption
{(a) Our empirically estimated redshift $z_a$ (given by 
Eqs.(\ref{eq:Lz3,1})-(\ref{eq:Hz2,2}))
versus the Sawicki et. al. (1997) template-fitting
photometric redshift $z_{temp}$, for 848 galaxies in the HDF
with $I<27$ and measured $UBVI$. 
The symbols are the same as in Fig.1.
The solid diagonal line indicates $z_a=z_{temp}$;
the dotted lines mark the region $|z_a-z_{temp}|\leq 0.5$.
(b) Same as Fig.2(a), but with the addition of
Eq.(\ref{eq:gap}) to select flat spectrum galaxies with
estimated redshifts given by Eq.(\ref{eq:zgap}). 
}
\end{figure}

There are 32 new spectroscopic redshifts of $z<4$ HDF galaxies
(Dickinson 1998, Phillips et al. 1998, Steidel et al. 1999) listed
in the recent paper by Fern\'andez-soto et al. (1999) which were
not available to us when we derived our photometric estimator
(Wang et al. 1998). Fig.3(a) shows the difference between our
estimated redshift $z_a$ and the spectroscopic redshift $z$,
scaled by $(1+z)$, as a function of $z$.
As can be seen, our empirical redshift estimator yields excellent
predictions for $z<2$, with $\sigma_z=0.13$ for all 18 $z<2$ galaxies,
and with $\sigma_z=0.08$ if we omit the most discrepant object.
For $z>2$ galaxies, we find $\sigma_z=0.34$ for all 14 galaxies, 
and $\sigma_z=0.31$ if we omit the galaxy with $z=2.002$.
The larger dispersion
between $z_a$ and $z$ for $z>2$ is largely due to the small number
of spectroscopic redshifts which were used in deriving the redshift
estimator.
\begin{figure}
\plottwo{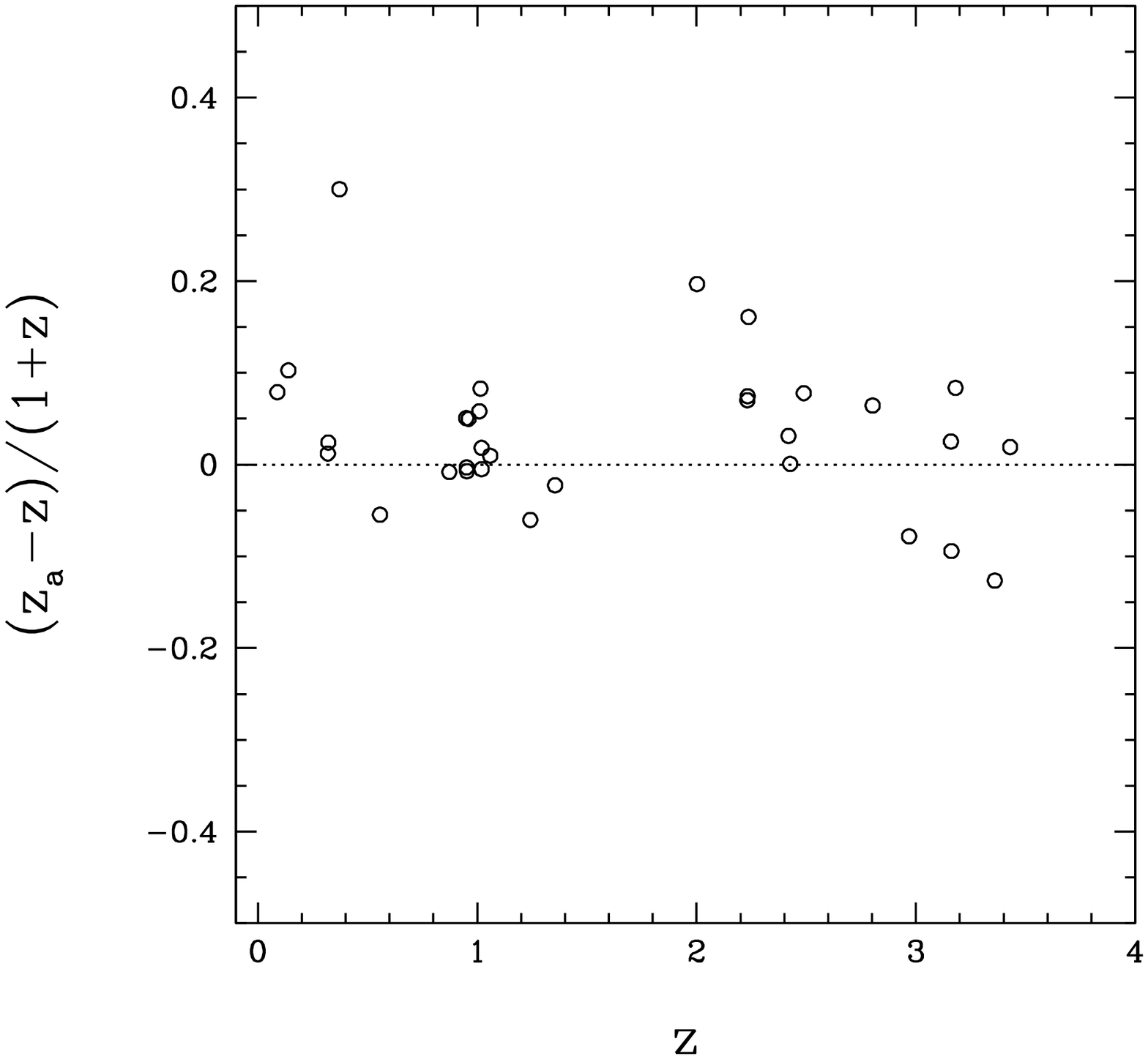}{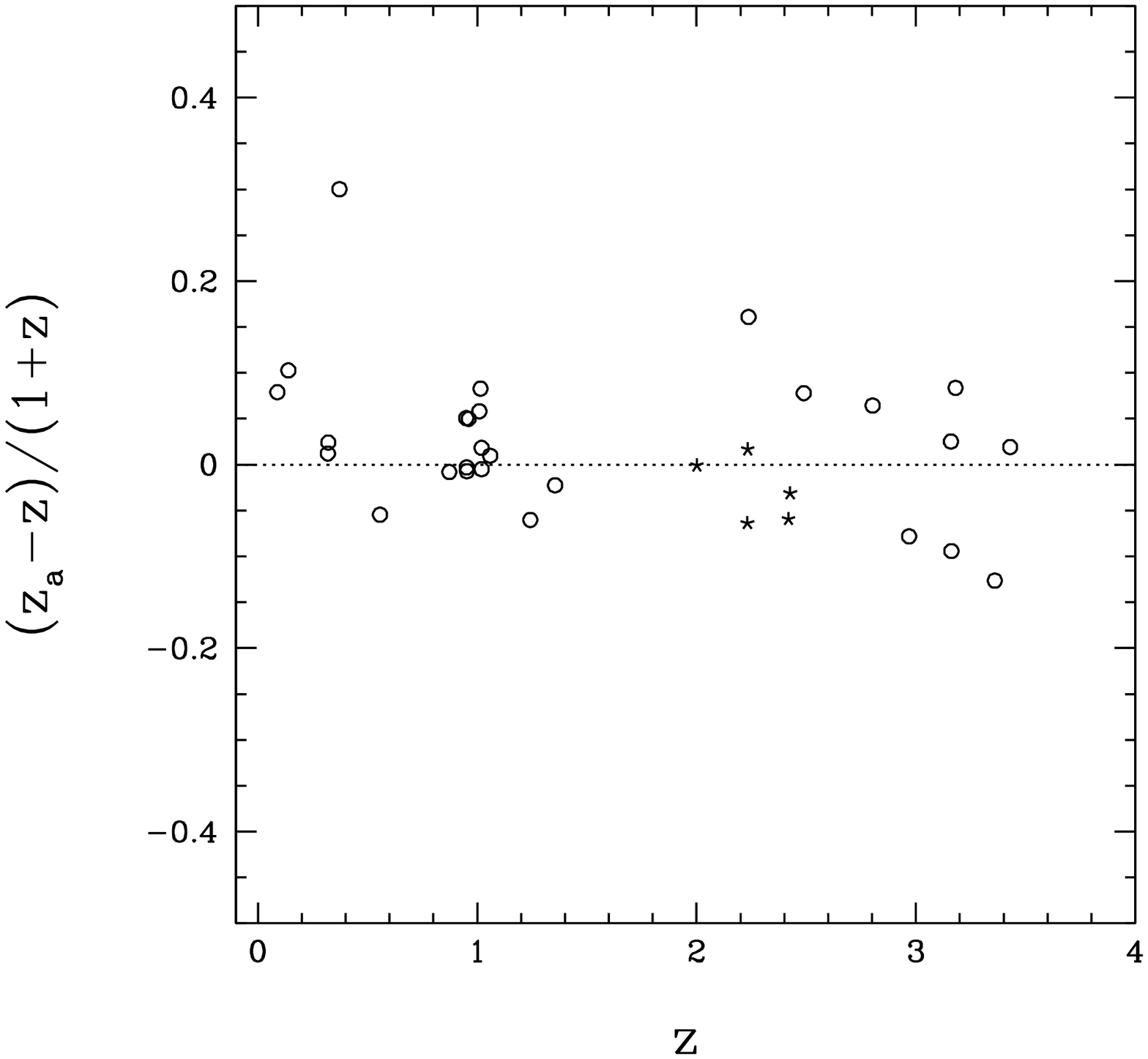}
\caption
{(a) The difference between our estimated redshift $z_a$ 
(given by Eqs.(\ref{eq:Lz3,1})-(\ref{eq:Hz2,2}))
and the spectroscopic redshift $z$,
scaled by $(1+z)$, as a function of $z$, for 32 new $z$'s.
(b) Same as (a), but with the addition of
Eq.(\ref{eq:gap}) to select flat spectrum galaxies with
estimated redshifts given by Eq.(\ref{eq:zgap}).
Star symbols represent the modified redshifts.}
\end{figure}

\section{Flat spectrum galaxies}

Galaxies in the redshift range $1.4\la z \la 2$ have flat spectra, 
which makes the measurement of their spectroscopic redshifts difficult.
The absence of calibrators in the range $1.4\le z\le 2.2$ (see Fig.1)
leads to an artificial gap in Fig.2(a) for $1.5 \la z_a \la 2$
in our estimated redshift $z_a$.

To illustrate how the gap in Fig.2(a) can be filled, we select galaxies
which satisfy
\ba
\label{eq:gap}
&& 23.4 \leq I<26.84, \, U\geq 26,\\ \nonumber
&&\left|V-I\right|\leq 0.6, \, \left|B-V\right|\leq 0.8, \,
0.6 \leq U-B \leq 2.
\ea
We assume that these galaxies satisfy the same color redshift 
relation as the color range cr$=3$ galaxies (which are closest 
to the intermediate redshift galaxies in color and redshift), but with 
a different constant offset (see Eq.(\ref{eq:Lz3,3})), i.e.,
\be
\label{eq:zgap}
z_a=1.5+0.480(U-B)-0.513(B-V)-0.250(V-I).
\ee
The constant offset value has been obtained by calibrating
with a galaxy which satisfies Eq.(\ref{eq:gap}) and has
spectroscopic redshift $z=2.002$. Fig.2(b) presents our estimated redshift
$z_a$ versus the Sawicki et al. template based photometric redshift
$z_{temp}$; it is the same as Fig.2(a), but with the addition of
Eq.(\ref{eq:gap}) to select flat spectrum galaxies (represented
by star symbols) with estimated redshifts given by Eq.(\ref{eq:zgap}). 
This addition improves on the previous comparison by moving
a number of galaxies with $z_{temp} \sim 2$ and
$z_a\ga 2.5$ or $z_a\la 0.5$ in Fig.2(a) into
the redshift gap of Fig.2(b). 

We can modify our empirical photometric redshift estimator by
adding Eq.(\ref{eq:gap}) to select flat spectrum galaxies with
estimated redshifts given by Eq.(\ref{eq:zgap}), before applying
Eqs.(\ref{eq:Lz3,1})-(\ref{eq:Hz2,2})).
Fig.3(b) shows the difference between our modified estimated redshift $z_a$ 
and the spectroscopic redshift $z$, scaled by $(1+z)$, as a function of $z$, 
for 32 newly measured galaxy redshifts. 
The modified redshift estimator 
makes the estimated redshifts for $z>2$ galaxies less biased, 
with $\sigma_z=0.30$ for 13 $z>2$ galaxies (excluding the calibrator galaxy 
with $z=2.002$).

\section{Discussion}

The advantage of our method is that it is model-independent and simple. 
The disadvantage of the method is that it requires a galaxy redshift sample
for training; this is clearly seen in Fig.2(a), where 
an artificial gap for $1.5 \la z_a \la 2$ exists as a result of
the absence of calibrators with measured redshifts 
in the range $1.4\le z\le 2.2$ (see Fig.1).
In this paper, we have illustrated how this gap in estimated redshifts
can be filled by modifying the empirical photometric redshift estimator
to include flat spectrum galaxies (see \S 3).
It is important to fill the gap in measured 
spectroscopic redshifts for $1.4\la z\la 2$,
as it would enable better calibration of photometric redshift estimators,
and better understanding of the nature of intermediate redshift galaxies.

Since our method is purely empirical, it is sensitive to the data properties
of the training set. In order to find and remove possible systematic effects 
in the photometric redshift estimator due to selection effects in the 
redshift measurements, it is important to have spectroscopic redshift 
surveys which are complete to a given magnitude limit.

The photometric redshift estimator described here is simple and accurate.
It can be further improved by additional data.

\acknowledgments
It is a pleasure to thank Ray Weymann for organizing this 
great workshop, and for encouraging Y.W. to attend it; and Howard
Yee for very helpful comments. 
We have used the Sawicki et al. (1997) HDF photometric catalog in this paper.

% Now comes the reference list.  Since we typed out the citations ourselves,
% the reference list is enclosed in a "references" environment.  Each
% new reference begins with a \reference command which sets up the proper
% indentation.  Typography that may be required in the reference list by
% the editorial staff must be included by the author.
%
% Observe the "standard" order for bibliographic material: author name(s),
% publication year, journal name, volume, and page number for articles.
% Some journal names are available as macros; see the WGAS markup
% instructions for a listing of which ones have been "macro-ized".
% Note the use of curly braces to delimit the font changes: it is essential
% that this be done to limit the scope of the font declaration.
%
% There is no need to engage in any other typographic manipulation.

\end{document}